\documentclass[%
 aip,
 amsmath,amssymb,
 reprint,
]{revtex4-1}
\usepackage{graphicx}% Include figure files
\usepackage[caption=false]{subfig} % <=== "caption=false" added
\usepackage[export]{adjustbox}

\usepackage{dcolumn}% Align table columns on decimal point
\usepackage{bm}% bold math
\usepackage{verbatim} %blok multi line comment
\usepackage{hyperref}% add hypertext capabilities
\hypersetup{
colorlinks=true,
linkcolor=blue,         
citecolor=blue,        
filecolor=blue,        
urlcolor=blue
}

\begin{document}

%-----------------------
\preprint{APS/123-QED}

\graphicspath{ {./figures/} } %look for all the images in 'images' folder

\title{Compact single-seed, module-based laser system on a transportable high-precision atomic gravimeter}

\author{F. E. Oon}
\email{fong\_en@nus.edu.sg}

\affiliation{Centre for Quantum Technologies, National University of Singapore, 3 Science Drive 2, 117543, Singapore}

 \author{Rainer Dumke}%
 
 \affiliation{Centre for Quantum Technologies, National University of Singapore, 3 Science Drive 2, 117543, Singapore}

\affiliation{Division of Physics and Applied Physics, Nanyang Technological University, 21 Nanyang Link, 637371, Singapore 
}%

\date{\today}% It is always \today, today,
             %  but any date may be explicitly specified
%---------------------------

\begin{abstract}
A single-seed, module-based compact laser system is demonstrated on a transportable $^{87}\text{Rb}$-based high-precision atomic gravimeter. All the required laser frequencies for the atom interferometry are provided by free-space acousto-optic modulators (AOMs) and resonant electro-optic phase modulators (EOMs). The optical phase-locked loop between the two optical paths derived from the same laser provides an easy frequency manipulation between two laser frequencies separated by the hyperfine frequency of 6.835 GHz using an AOM and an EOM, respectively. Our scheme avoids parasite Raman transitions present in the direct EOM modulation scheme (modulating directly at the frequency of the hyperfine splitting), which have detrimental effects on the accuracy of the gravity measurements. The optical phase-locked loop also provides a convenient way for vibration compensation through the Raman lasers' phase offset. Furthermore, the modular design approach allows plug-and-play nature on each individual optic module and also increases the mechanical stability of the optical systems. We demonstrate high-precision gravity measurements with 17.8 $\mu\text{Gal}$ stability over 250 seconds averaging time and 2.5 $\mu\text{Gal}$ stability over 2 h averaging time.  

\end{abstract}

\maketitle

\section{
\label{sec:level1}
Introduction}

Atomic (quantum) gravimeters measure acceleration of free-falling atomic test masses by using matter-wave interferometry~\cite{peters2001high,geiger2020high}. Such gravimeters have demonstrated comparable performance to its classical counterpart (absolute gravimeter using free-falling corner cube~\cite{niebauer1995new}) in terms of sensitivity, long-term stability and accuracy~\cite{freier2016mobile,gillot2014stability}. Due to its ability to carry out continuous and long-term absolute gravity measurements (not possible by free-falling corner cube gravimeters due to mechanical wear during each drop of measurement), great development efforts are currently being dedicated to transform atomic gravimeters from  lab-based instruments into field-capable or on-board devices~\cite{menoret2018gravity,bidel2018absolute}.
\par

The laser system for alkali-based atomic gravimeters usually requires at least two master lasers in order to address the atomic hyperfine transitions that are few GHz apart~\cite{hu2013demonstration,menoret2011dual,leveque2014laser,zhang2018compact}. The master laser remains one of the most delicate systems within atomic interferometry experiments, especially when implementing vibration compensation via an optical phase-locked loop of the two Raman lasers~\cite{merlet2009operating, lautier2014hybridizing,le2008limits}. Hence, reducing the number of master lasers would greatly reduce the complexity of the experiments. A fiber-based single-seed laser system operating at $1.5$ $\mu\text{m}$ wavelength and down converted to $780$ nm wavelength using frequency doubling (periodically poled lithium niobate crystal) has been realized~\cite{theron2015narrow}. Single-seed laser systems operating directly at the desired wavelength for $^{85}\text{Rb}$ atoms~\cite{fang2018realization} and Cesium atoms~\cite{wu2017multiaxis} have also been developed. All of the above-mentioned experiments using single-seed laser generate Raman frequency pairs by directly modulating the EOM at the frequency of the hyperfine splitting. The infinite pairs of EOM carrier and sideband frequencies having the same frequency difference inevitably induce parasite Raman transitions that reduce the accuracy of the gravimeter measurements~\cite{carraz2012phase}. 
\par 

In this work, we demonstrate a module-based free-space laser system with a single $780$ nm wavelength master laser, addressing $^{87}$Rb atoms and offer flexibility in replacing/re-designing the optical systems. The cooling and repumper laser frequencies are generated by modulating the EOM at $\approx$ 6.58 GHz. The +1 sideband of the EOM modulation also serves as the higher-frequency Raman laser, while the lower-frequency Raman laser is generated by a double-pass acousto-optic modulator (AOM) having detuning of $-2\times 123$ MHz. This configuration results in a sole pair of Raman lasers with frequency separation of $\approx$ 6.83 GHz, thus avoiding the parasite Raman transitions such as in~\cite{theron2015narrow,fang2018realization,wu2017multiaxis}. The optical phase-locked system between the two Raman lasers offers a feedback mechanism for vibration compensation via the Raman lasers' phase offset. It also provides an easier Raman frequency manipulation at low frequency carrier (via AOM at 123 MHz) instead of manipulation at high carrier frequency of 6.83 GHz.

\section{Laser System}

\subsection{Optical distribution}

\begin{figure}

\subfloat{%
  \includegraphics[width=0.49\textwidth]{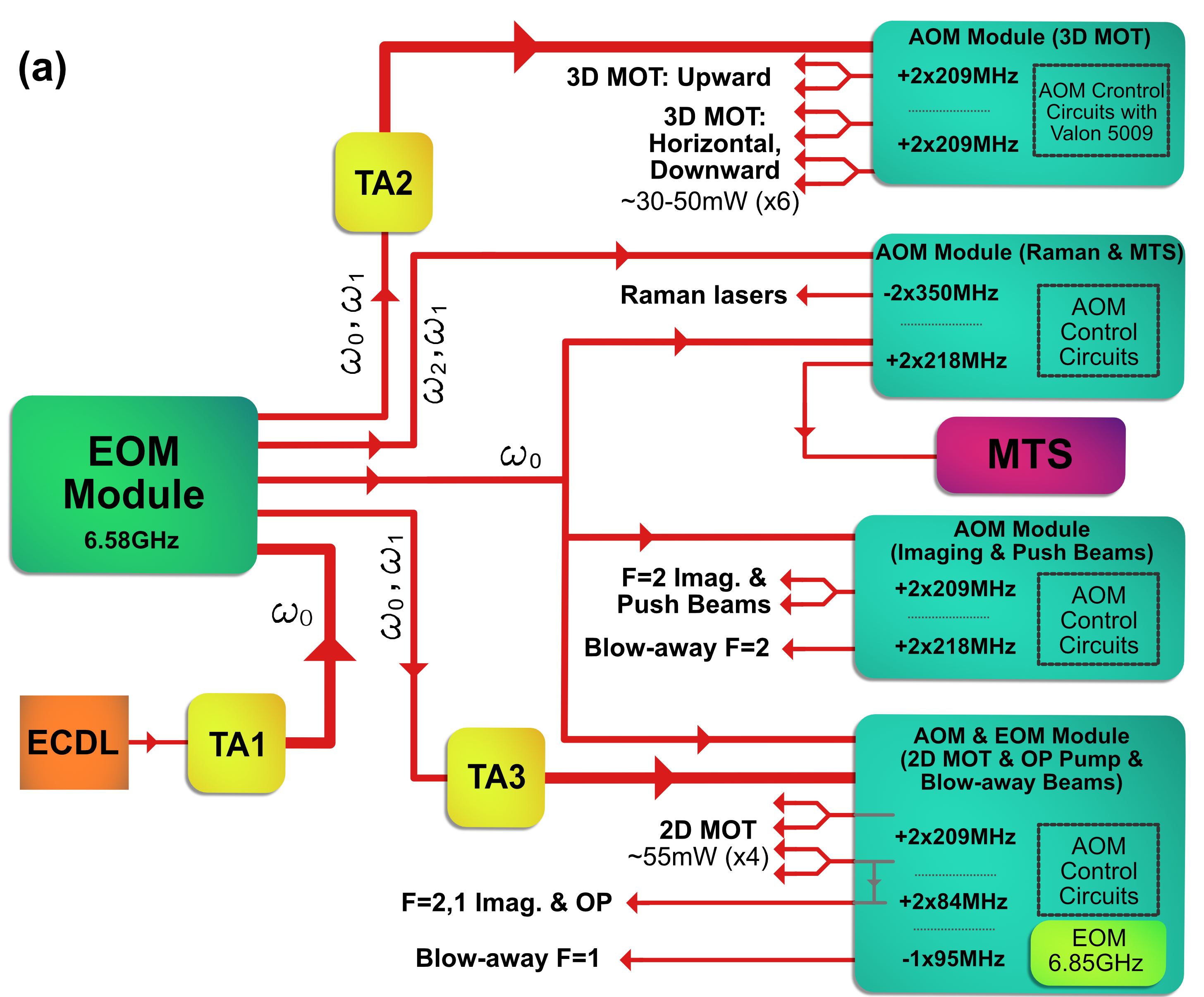}\label{subfig:OpticalPaths}%
  
}\hfill
\subfloat{%
  \includegraphics[width=0.49\textwidth]{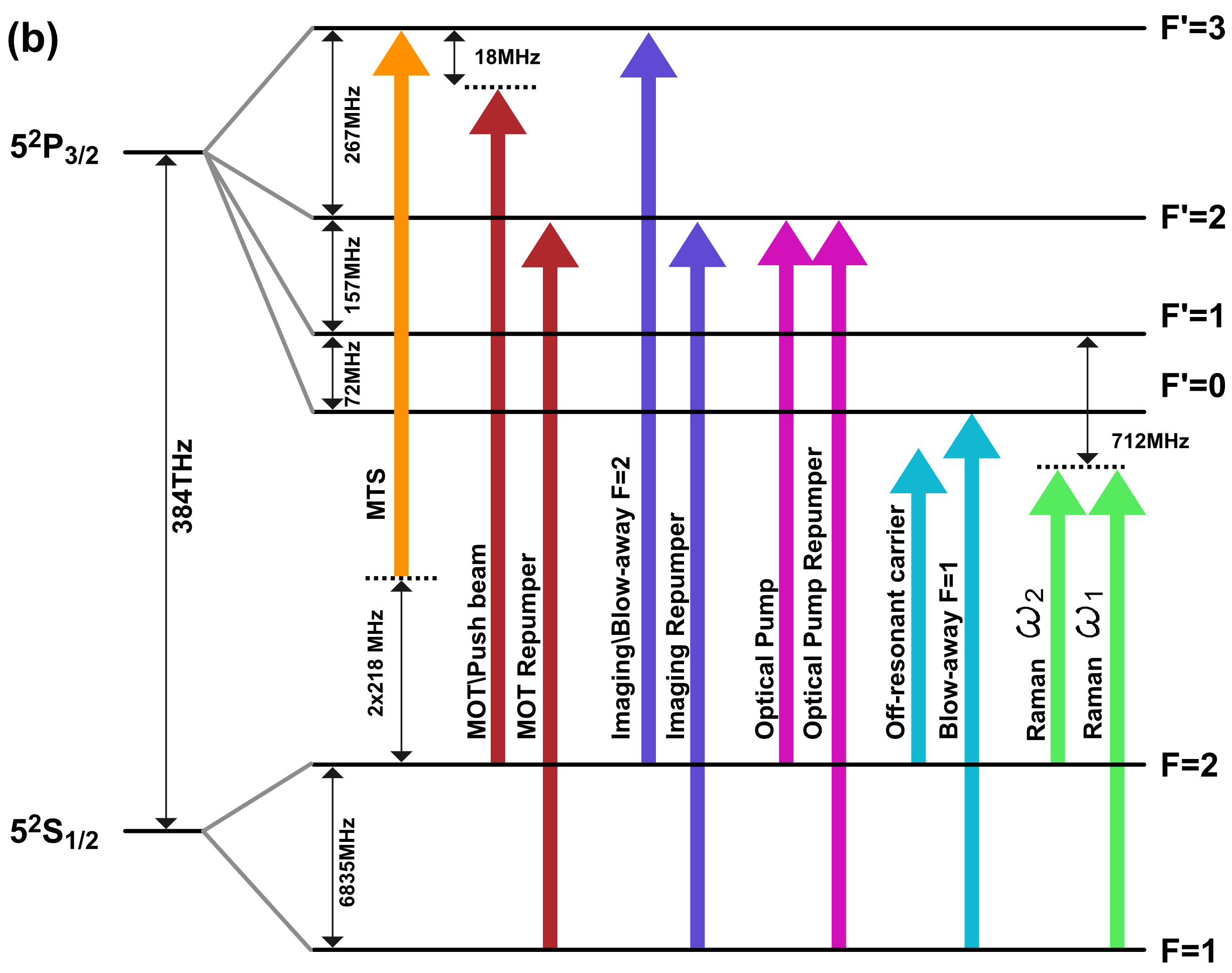}%
  \label{subfig:FrequencyTransitions}
}\hfill
  
\caption[]{(a) Optical distribution system. Each rectangular box represents an independent optic module. Red arrows represent optical fiber connections. All laser frequencies are derived from a single external-cavity diode laser (ECDL). The optical output of the ECDL is amplified by a tapered (optical) amplifier (TA), resulting in $\approx$ 1 Watt of power after coupling into the polarization maintaining fiber. The amplified laser (unmodulated, $\omega_0$) is then coupled into a main optic module where one of the Raman lasers ($\omega_1$) and the repumper laser (for magneto-optical trap) are generated by the +1 sideband of a resonant electro-optic phase modulator (EOM, resonant at 6.58 GHz) while the second Raman laser ($\omega_2$) is generated by a double-pass acousto optic modulator (AOM). The lasers are then further distributed to different optic modules, generating the required frequencies via another 8 AOMs and 1 EOM (resonant at 6.85 GHz) - here the +1 sideband corresponds to the repumper for the optical pumping process, as well as blow-away light for the atomic transition $F=1\leftrightarrow F'=0$. (b) Atomic energy levels and laser frequencies. All the laser frequencies are referenced to the transition $F=2\leftrightarrow F'=3$ via modulation transfer spectroscopy (MTS, as described in~\cite{mccarron2008modulation}). Each color pair represents the carrier frequency and the $+1$ sideband of the EOMs.     }
\label{fig:OpticDistributions}
\end{figure}

We place all of the electronic and optical components that serve the same function into separate aluminum enclosures (modules), see Figures~\ref{fig:OpticDistributions} and~\ref{fig:Hardware}. For example, the temperature and current controllers for the master laser are grouped together with the laser itself into a single module. Different modules are interconnected via polarization-maintaining optical fibers. This modular arrangement results in stand-alone packages, such as the master laser module, the optical amplifier module and the spectroscopy module. The modular design allows for re-arrangement of the optical paths, as well as modification to individual modules without affecting the others, providing a high degree of re-configurability. 
\par
All the laser frequencies are derived from a single interference-filter-stabilized external-cavity diode laser (ECDL, linewidth $<$100 kHz) module ~\cite{baillard2006interference,gilowski2007narrow}. The influence of the master laser phase noise on the sensitivity of retro-reflected atom interferometers can be found in~\cite{le2007influence}. Three tapered (optical) amplifier (TA) modules~\cite{zappala2014note} are employed to
obtain the desired laser intensity (Thorlabs TPA780P20). Different laser frequencies are provided by two resonant electro-optic phase modulators (EOMs; QUBIG GmbH) and nine acousto-optic modulators (AOM; AA Opto Electronic). 

%electrical cables

\subsection{Module-based design}

\begin{figure}

\subfloat{%
\centering
  \includegraphics[width=0.4\textwidth]{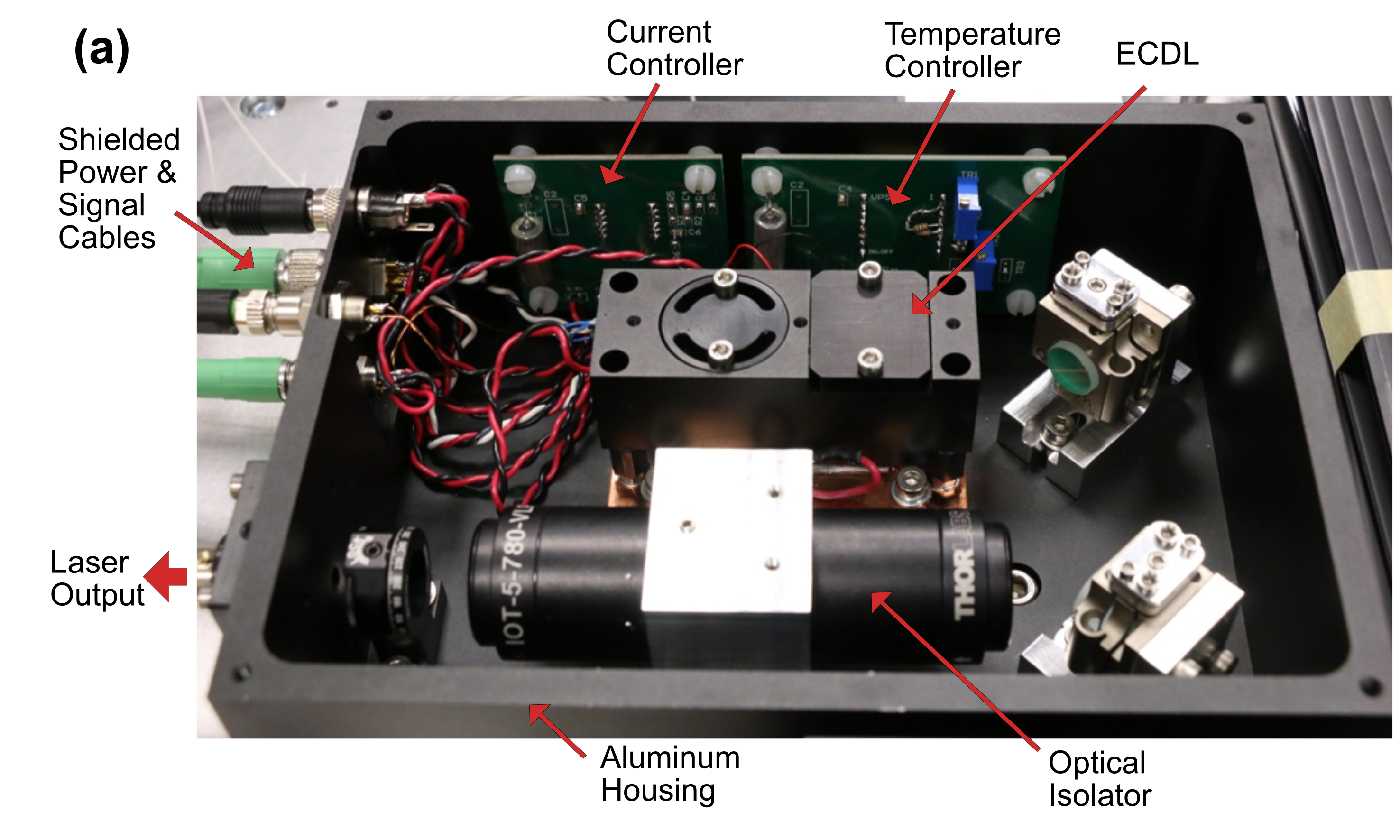}%
  \label{subfig:LaserModule}
  }\hfill
  
\subfloat{%
\centering
  \includegraphics[width=0.4\textwidth]{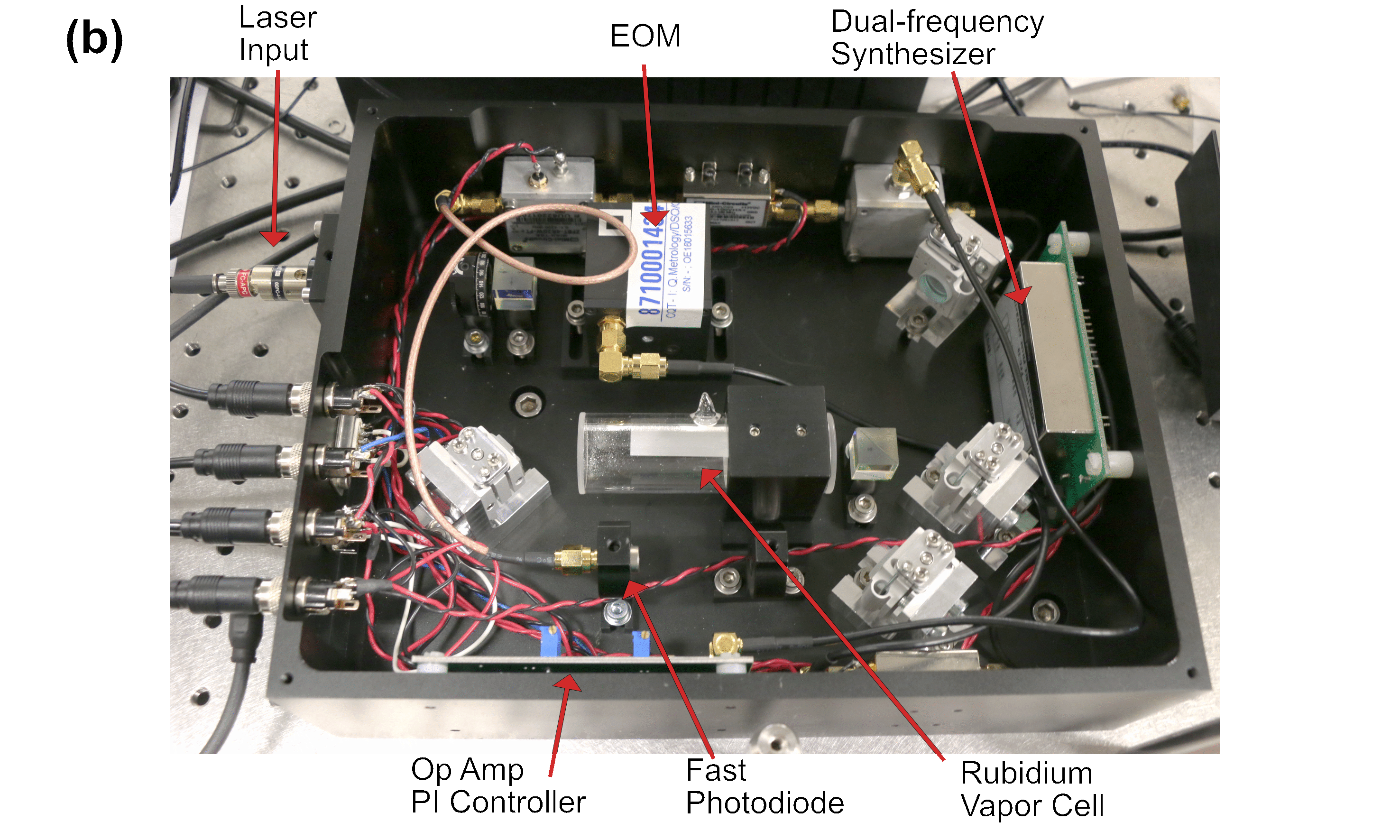}%
  \label{subfig:MTSModule}
}\hfill

\subfloat{%
\centering
  \includegraphics[width=0.4\textwidth]{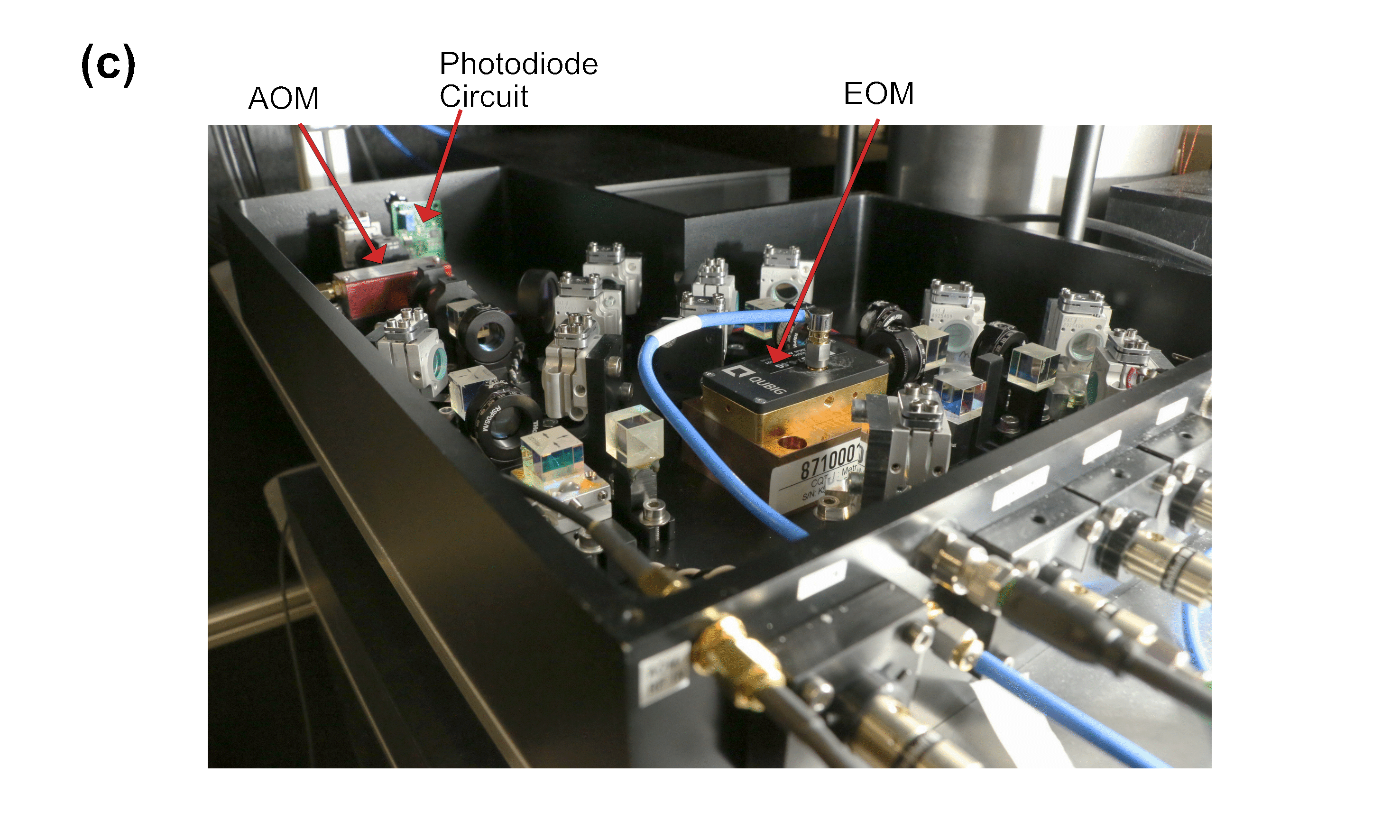}%
  \label{subfig:EOMModuleActual}
}\hfill

\subfloat{%
\centering
  \includegraphics[width=0.4\textwidth]{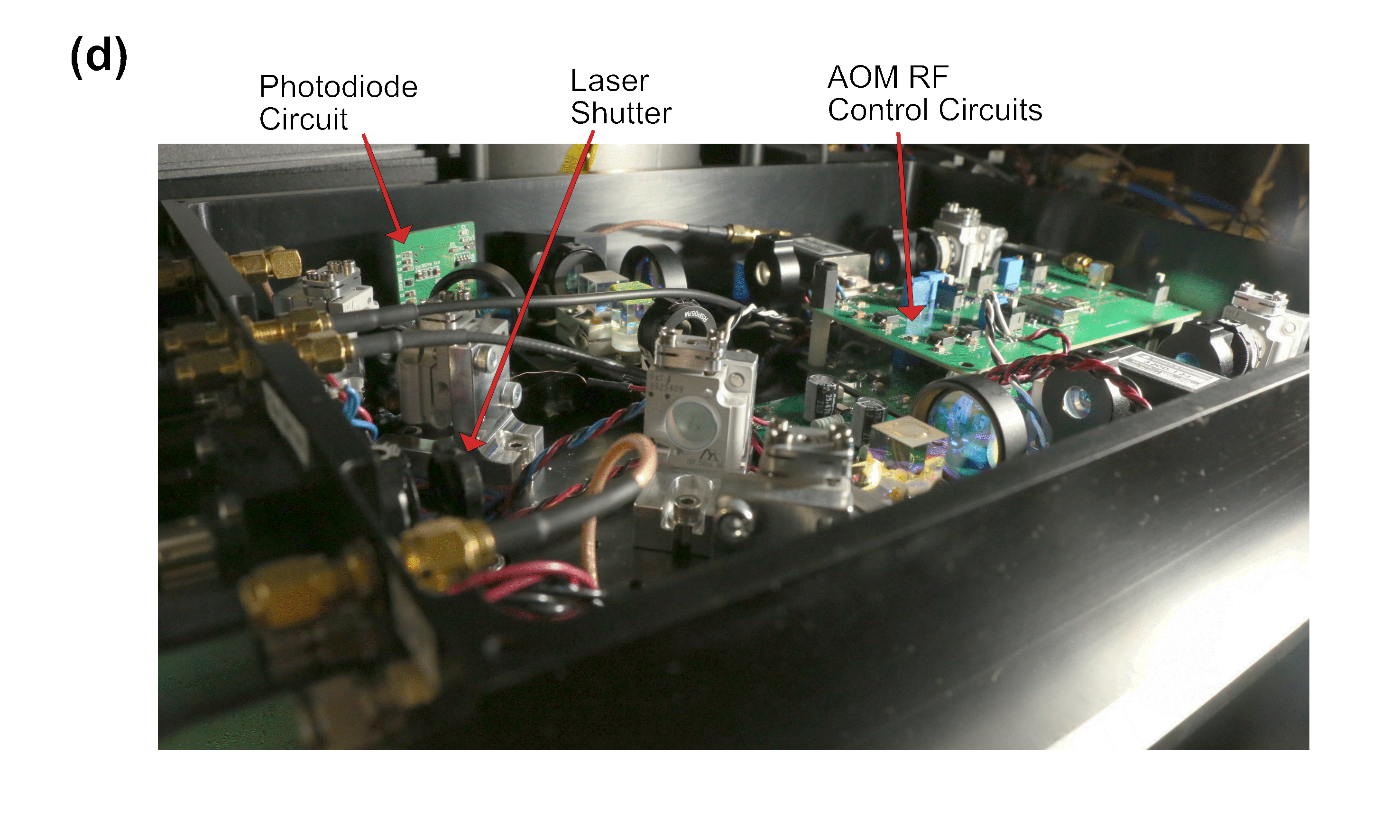}%
  \label{subfig:AOMModule}
}\hfill
  
\caption[]{Actual photos of the optic modules with covers removed: (a) ECDL module, (b) MTS module, (c) EOM module and (d) AOM module. Each independent module is optically connected by the compact fiber couplers (Schäfter+Kirchhoff GmbH, 60FC) which allow re-connection of optical fibers without affecting the optical alignments of the optic modules. Electronic components that generate moderate heat are attached to the aluminum wall to ensure good heat dissipation. }
\label{fig:Hardware}
\end{figure}
 
\begin{figure}%[h]
%\centering
\includegraphics[width=0.48\textwidth]{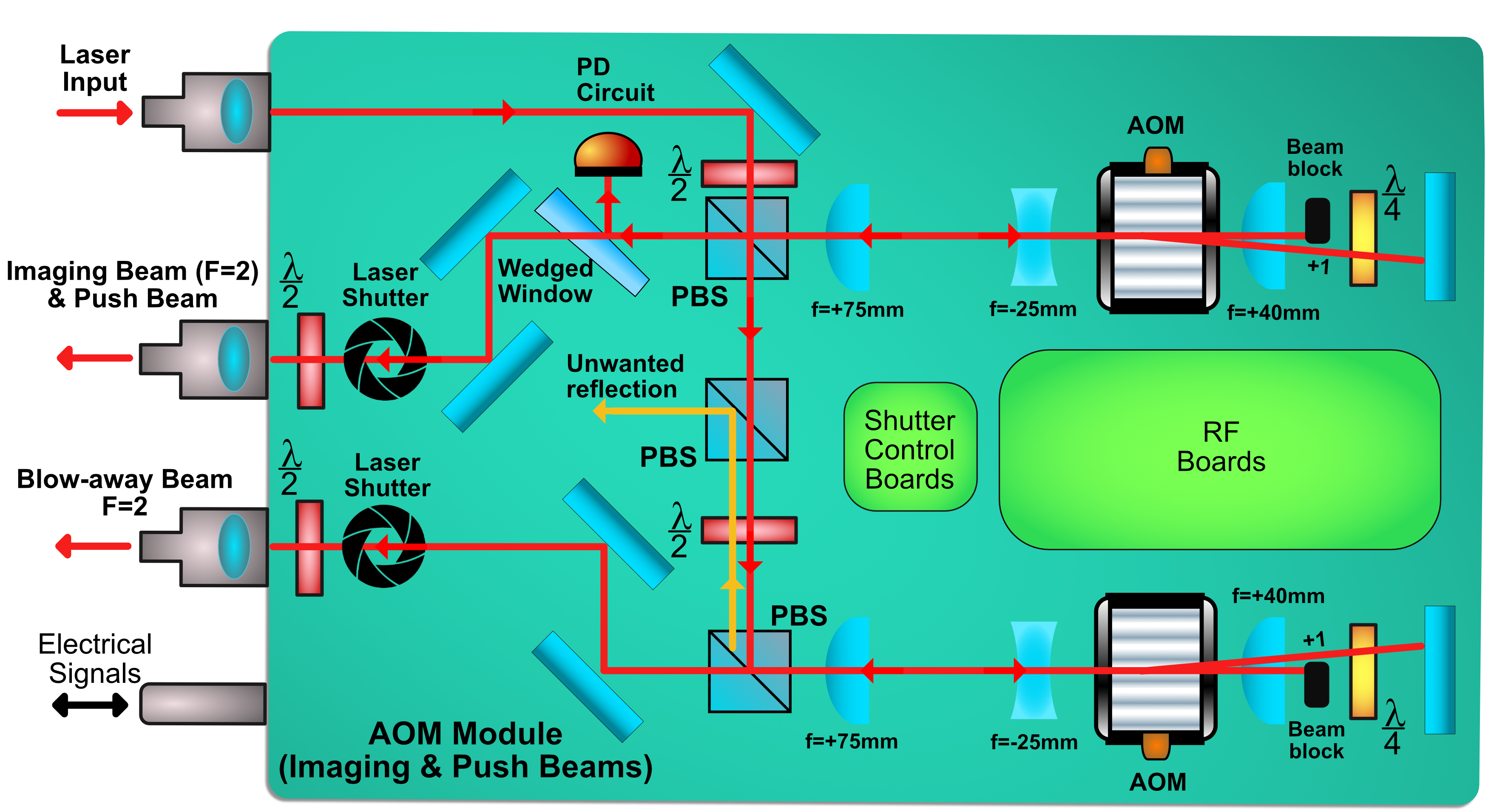} 
\caption[]{AOM module for the imaging and push beams. A small amount of light is reflected into a self-made photodiode circuit. The signal is then feedback to a voltage-controlled RF attenuator to stabilize the laser intensity to $<$ 0.5 \% by varying the diffraction efficiency of the double-pass AOM.} 
\label{fig:ImagingAOMSchematic} 
\end{figure}

Figure~\ref{fig:Hardware} shows the actual photos of our atomic gravimeter's optic modules. Each module's enclosure is cut out from a single piece of aluminum, with total height of $73$ mm, base thickness of $10$ mm and laser beam height maintained at $30$ mm. The modules are designed to be air-tight to prevent air disturbances and accumulation of dust. To ensure long-term mechanical stability of the laser beams, we employ flexure type mirror holders (Siskiyou, IXF.50ta) and compact fiber couplers (Schäfter+Kirchhoff GmbH, 60FC). The mechanical stability of the laser beams are further improved due to the fact that the optical paths are truncated into individual modules interconnected by optical fibers. Each module is controlled externally by a computer via shielded electrical cables. We showcase the schematic design of the imaging beam and push beam pertaining to the AOM module in Figure~\ref{fig:ImagingAOMSchematic}. We have not observed misalignment of the optical elements by transporting the gravimeter within the building.

\section{Frequency control}
\subsection{EOM module}

\begin{figure}

\subfloat{%
  \includegraphics[width=0.49\textwidth]{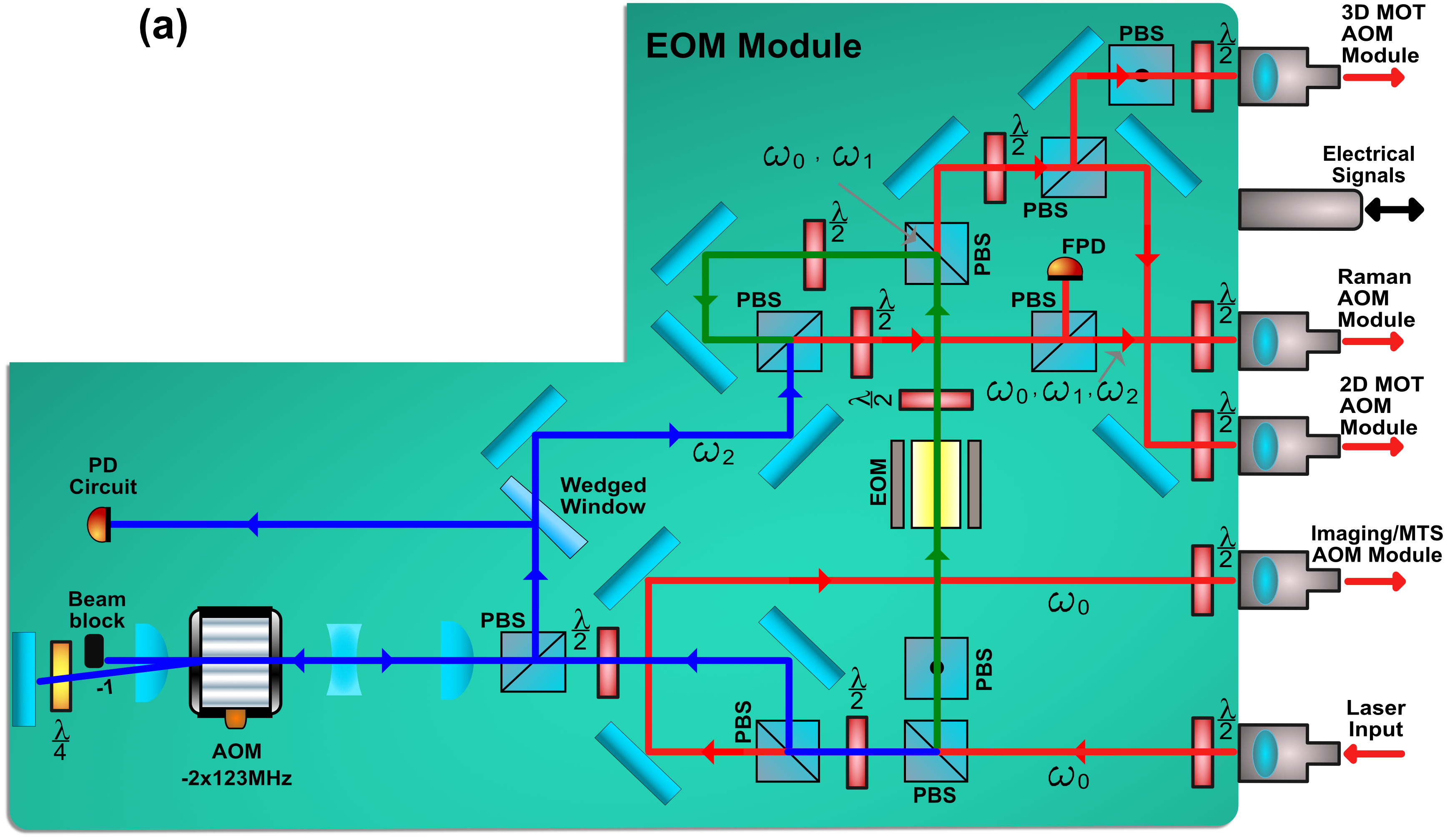}%
  \label{subfig:EOMModuleSchematic}
}\hfill
\subfloat{%
  \includegraphics[width=0.4\textwidth]{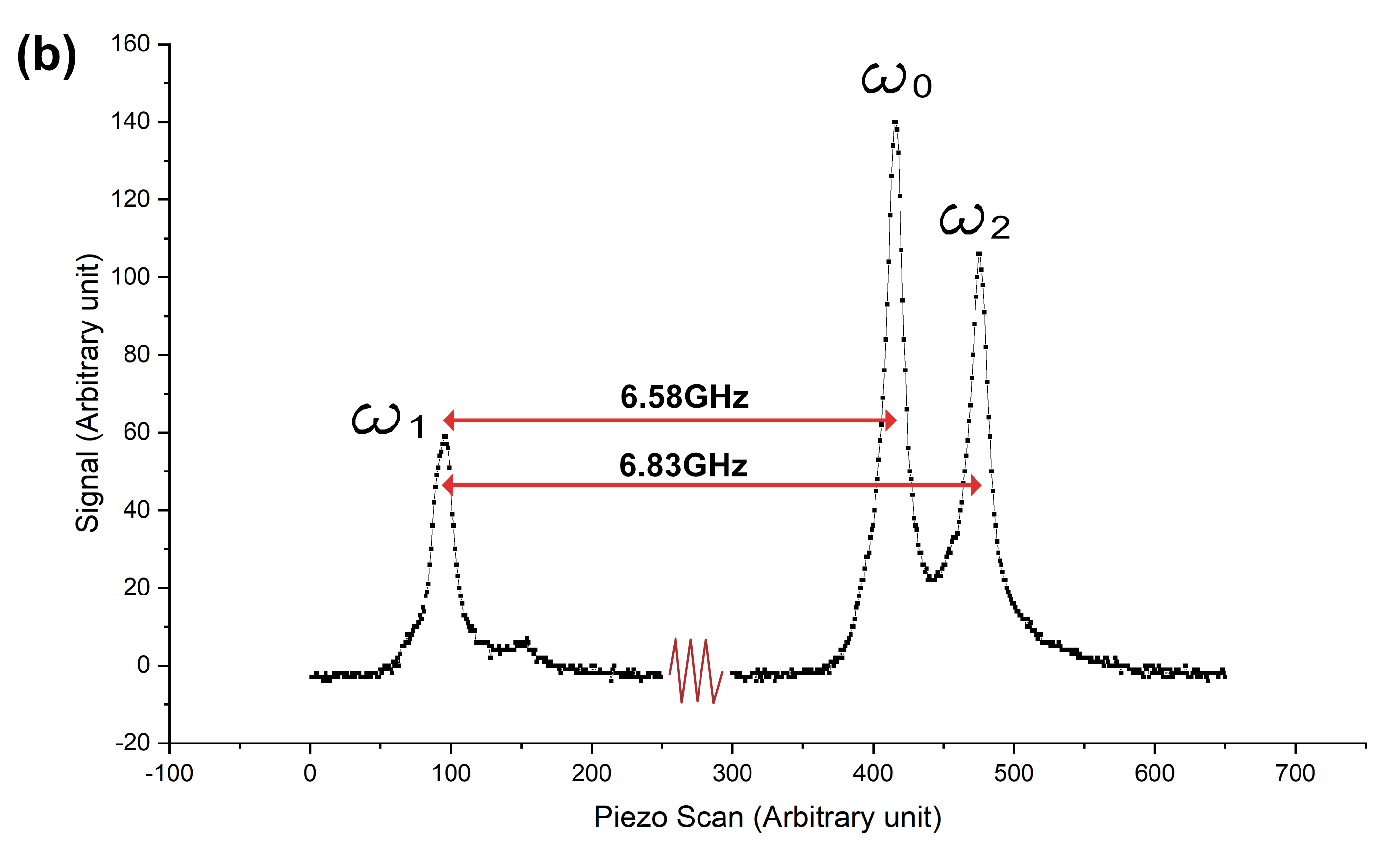}%
  \label{subfig:RamanEOMFP}
}\hfill
  
\caption[]{(a) Optical layout of the EOM module: A portion of the un-modulated laser beam ($\omega_0$) is diverted to an EOM resonant at $6.58$ GHz (green path). The $+1$ sideband serves both as the repumper light for the magneto-optical trap (+1 sideband to carrier intensity ratio: 2\%) and one of the Raman lasers (higher frequency, $\omega_1$, +1 sideband to carrier intensity ratio: $>40$ \%, equivalent to $\approx$ 2.5 Watts of RF power for the resonant EOM). A second portion of the un-modulated laser beam (blue path) goes through a double-pass AOM ($-2\times 123$ MHz), where the lower frequency side of the Raman laser ($\omega_2$) is generated, resulting in the hyperfine splitting frequency of $\omega_1-\omega_2=6.58+2\times 0.123\approx 6.83\text{GHz}$. $\omega_1$ and $\omega_2$ are overlapped into same polarization before coupling into an optical fiber. The PBS symbol with a dot in the center indicates s-polarization with upward orientation. (b) Fabry-Pérot cavity spectrum of the laser beams $\omega_0$, $\omega_1$ and $\omega_2$.}
\label{fig:EOM}
\end{figure}

During the free fall of the atomic test masses, in this case rubidium-87 atoms, the acceleration of the atoms is interrogated by the counter-propagating Raman lasers. The frequency difference between the two counter-propagating Raman lasers ($\approx$ 6.8 GHz) are chirped to compensate the induced Doppler shift of the falling atoms due to gravitational acceleration, which is around $25.07$ MHz/s at 1 g. This chirping rate is served as the ``optical ruler'' of the gravity measurements where the gravitational acceleration can be approximated as $g\approx\frac{2\pi}{k_{\text{eff}}}\frac{df}{dt}$~\cite{peters2001high}, where $\frac{df}{dt}$ is the frequency chirping rate and $\frac{2\pi}{k_{\text{eff}}}\approx \frac{780}{2}$ nm is the effective wavelength of the Raman lasers. 
\par
In order to attain a measurement precision of $\frac{\Delta\text{g}}{\text{g}}=10^{-9}$, the frequency precision of the Raman lasers' driving frequency has to be maintained at least at the same level, $\frac{\Delta\text{f}}{\text{f}_0}=10^{-9}$~\cite{freier2017atom}, where $f_0=6.8$ GHz. Direct chirping of $6.8$ GHz frequency at sub-Hz precision for a duration of $\approx 1$ s can be technically challenging. In this work, the Raman laser frequency difference is driven by two separate frequency synthesizers, namely, a synthesizer (NI, QuickSyn Lite Synthesizer) driving the high frequency side of the Raman lasers ($\omega_1$, 6.58 GHz) and a direct digital synthesizer (DDS, AD9854) driving the low frequency side of the Raman lasers ($\omega_2$, 123 MHz), see Figure~\ref{fig:EOM}. The frequency chirping is provided by the AD9854 via a double-pass AOM during the interferometry sequence, ensuring deterministic frequency shifting and micro-second time steps~\cite{karcher2020impact}. The intensity ratio between the Raman lasers is easily adjusted via the half-wave plate and polarization beamsplitter pair (which overlaps the two laser beams) in order to minimize the differential AC Stark shift of the laser transitions~\cite{olbertz2011mobile}.

\subsection{Optical phase-locked loop and frequency chain}

\begin{figure}

\subfloat{%
  \includegraphics[width=0.98\columnwidth]{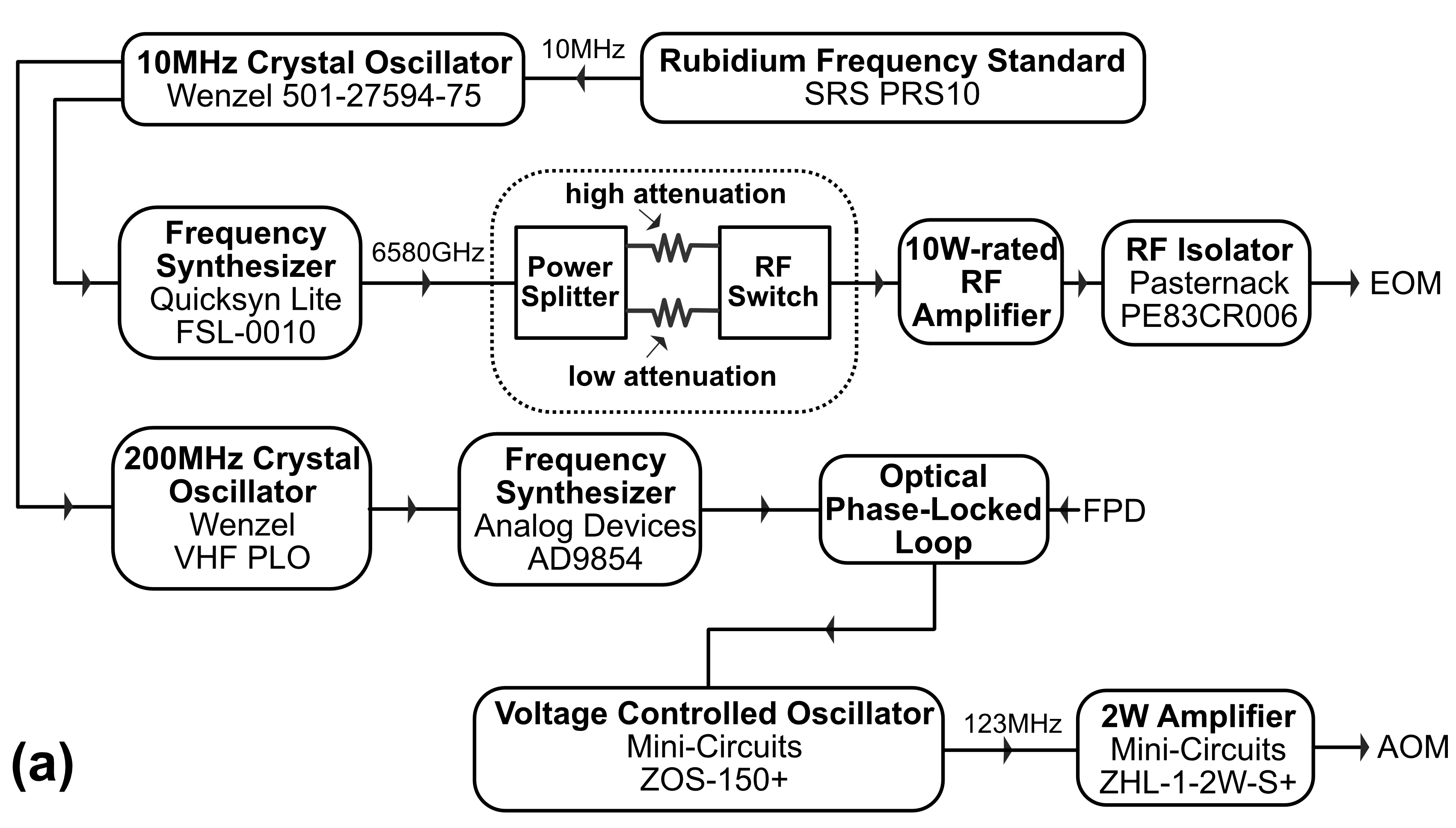}%
  \label{subfig:frequencyChain}
  
}\hfill
\subfloat{%
  \includegraphics[width=0.98\columnwidth]{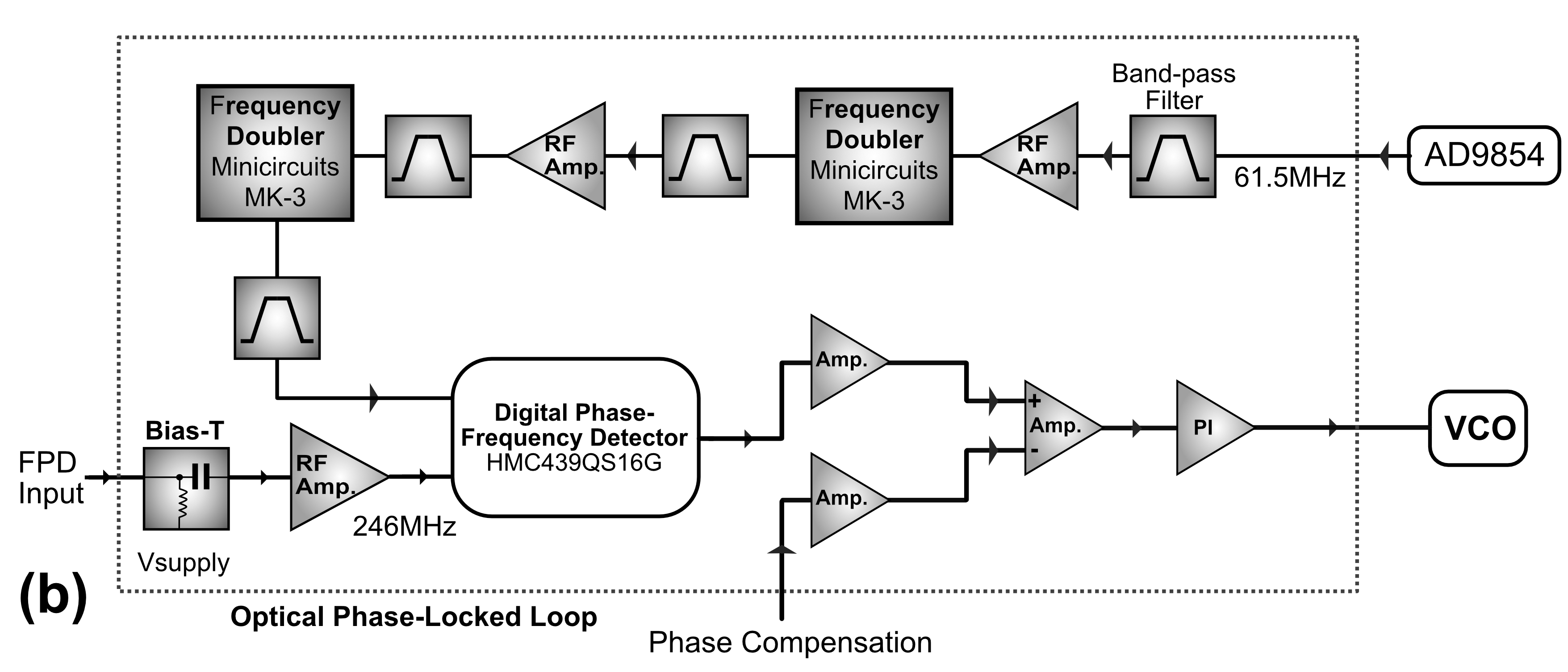}%
  \label{subfig:phaseLockedLoop}
}\hfill

\subfloat{%
  \includegraphics[width=0.98\columnwidth]{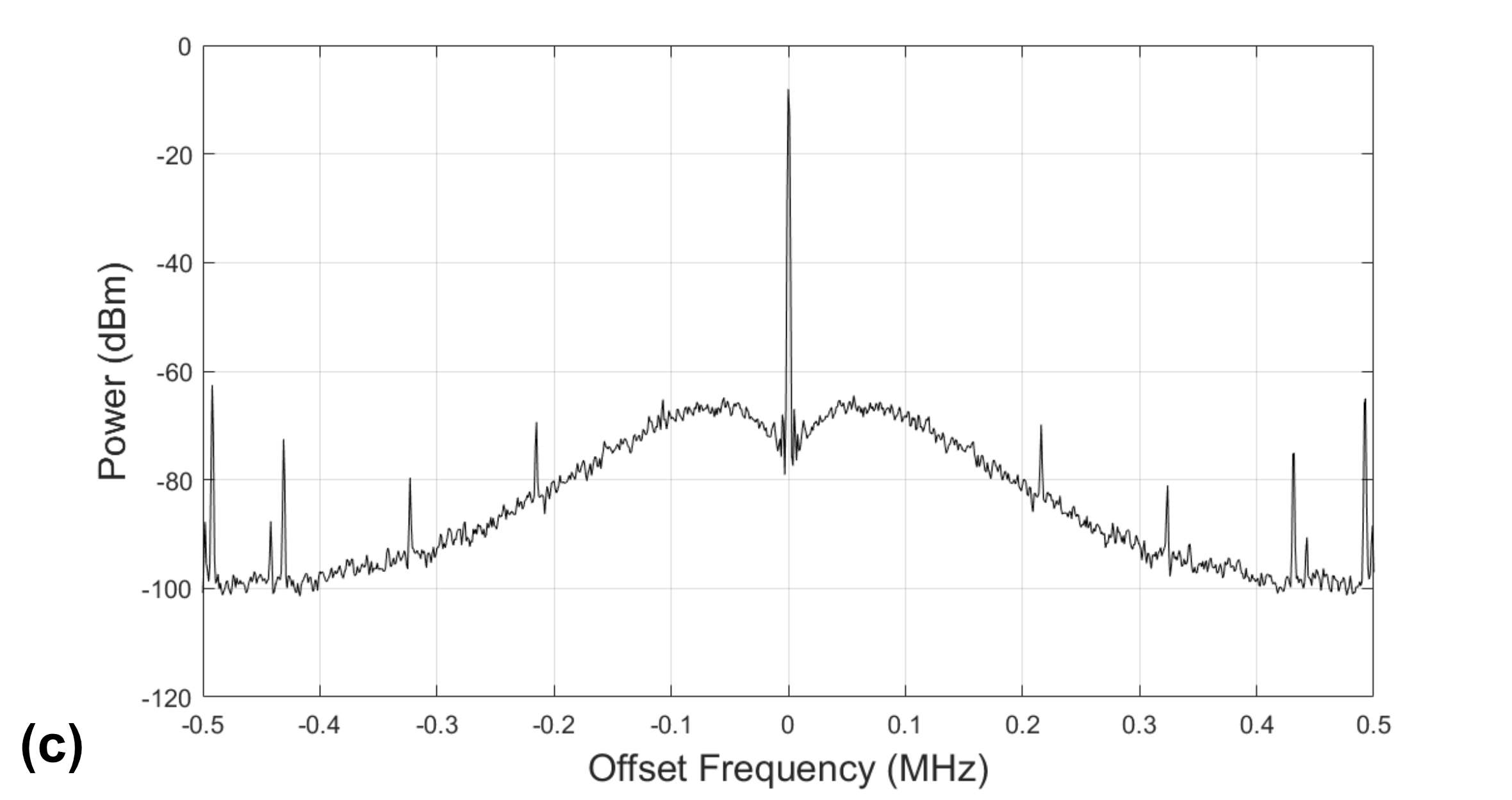}%
  \label{subfig:beatSignal}
}\hfill
  
\caption[]{(a) Frequency chain of the Raman lasers. All the laser frequencies are referenced to a 10 MHz Rubidium frequency standard, with an additional 10 MHz quartz oscillator for improved phase noise performance. Dashed box can be replaced by a single voltage-controlled attenuator (Mini-Circuits, ZX73-123+). (b) Schematic of the optical phase-locked loop. Frequency output of the AD9854 is multiplied by two frequency doublers before being compared with the beat signal from $\omega_0$ and $\omega_2$. The error signal is then used to lock the frequency of the voltage controlled oscillator (VCO, Mini-Circuits, ZOS-150+) via an operational amplifier proportional integrator (PI). (c) Beat spectrum (centered at 240 MHz) after phase lock, with PI servo bumps of $\approx$ 100 kHz wide. }
\label{fig:frequencyChain}
\end{figure}

All the frequencies in the gravimeter are referenced to a 10 MHz rubidium frequency standard (Stanford Research Systems, PRS10), Figure~\ref{subfig:frequencyChain}. Because the Raman lasers $\omega_1$ and $\omega_2$ travel two separate paths, an unwanted phase shift can be induced by mechanical vibrations, thermal drifts and air current disturbances. To maintain a fixed phase difference between the Raman lasers during the interferometry sequences, the beat frequency between $\omega_0$ and $\omega_2$ ($\approx 246$ MHz) after being recombined into the same polarization is immediately recorded by a fast photodiode (Hamamatsu Photonics, G4176-03) and then phase locked with the frequency output of the AD9854, Figure~\ref{subfig:phaseLockedLoop}. This enforces a phase locked condition between $\omega_1$ and $\omega_2$ since $\omega_0$ and $\omega_1$ travel the same path. The optical phase-locked loop also provides a convenient feedback path for vibration compensation through Raman lasers' phase offset (Section~\ref{subsec:phaseCompensation}). Because both of the Raman lasers are derived from a single ECDL, the optical phase-locked loop only requires a single slow feedback path (instead of an additional fast feedback path for wider locking bandwidth~\cite{yim2014optical}), Figure~\ref{subfig:beatSignal}. 
\par 

The overall frequency of the Raman lasers is further detuned by a double-pass AOM with $-2\times 350$ MHz detuning, resulting in a laser power of approximately 18 mW for $\omega_2$ and 9 mW for $\omega_1$ into the atoms. Together with a collimated Raman laser beam diameter of $\approx$ 20 mm, we have a $\pi-$pulse duration $=$ 34 $\mu\text{s}$. The authors would like to bring the attention of the readers that amplifying the Raman lasers intensity by seeding it into another TA would generate unwanted sidebands due to intensity modulation between the two close laser frequencies $\omega_0$ and $\omega_2$~\cite{ferrari1999high}. This is not a problem for the cooling laser in which $\omega_0$ and $\omega_1$ are $\approx 6.5$ GHz apart. If additional laser power is required for the Raman lasers, the carrier $\omega_0$ can be filtered out by a Fabry-Pérot cavity before overlapping with $\omega_2$ and subsequently amplified by the TA, as was carried out by~\cite{li2017phase}. Alternatively, the carrier can be nulled by increasing the power of EOM's RF amplifier. All of these methods would require phase locking at 6.8 GHz instead of $2\times 123$ MHz as is carried out by this work.

\subsection{Timing sequences}

\begin{figure}%[h]
%\centering
\includegraphics[width=0.48\textwidth]{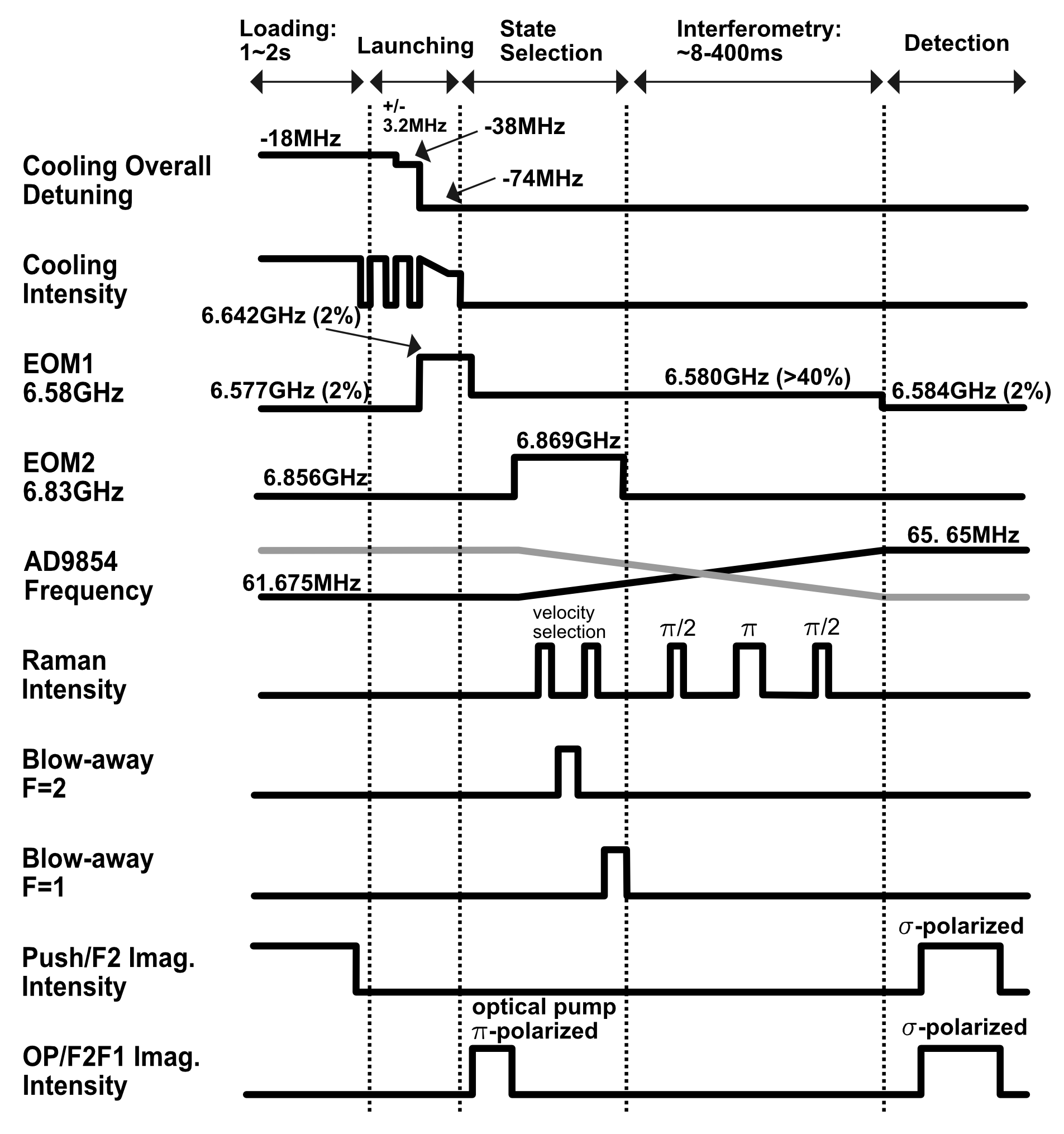} 
\caption[]{Timing sequences of the various laser frequencies.} 
\label{fig:TimingSequences} 
\end{figure}

The driving frequencies and the driving amplitudes of the EOMs need to be adjusted during different stages of the experiment sequence (Figure~\ref{fig:TimingSequences}). These are especially crucial for avoiding un-intended atomic transitions due to the multi-frequencies EOM's output. We briefly discuss the procedure as follow:    

\begin{enumerate}
    \item \textbf{Atom loading}: Only a small amount of repumper laser intensity is needed during magneto-optical trap (MOT) loading stage. Therefore, the driving amplitude of the 6.58 GHz-resonant EOM (EOM1) is lowered down such that +1 sideband to carrier intensity ratio is $\approx 2$\%. We typically load $\approx 0.5\sim 1\times 10^8$ atoms into the 3D MOT in $1\sim 2$ s.  
    \item \textbf{Launching}: The upward traveling cooling lasers are blue-detuned 3.2 MHz while the downward traveling cooling lasers are red-detuned 3.2 MHz for about 1.5 ms so that the atomic cloud is launched upwards to form an atomic fountain~\cite{peters2001high} trajectory (detailed experiment procedures please refer to~\cite{en2021transportable}). By decreasing the overall cooling laser frequency and intensity (far-detuned molasses~\cite{salomon1990laser}), the atomic cloud is further cooled to $\approx$ 3 $\mu\text{K}$. At the end of the launching procedure, EOM1's driving frequency is shifted to 6.642 GHz, so that the Repumper laser (+1 sideband) is resonant with $F=1\leftrightarrow F'=2$ while the cooling laser (carrier) is off-resonant to the atomic transition. This is to ensure that the atoms are pumped into F=2 hyperfine state.  
    \item \textbf{State selection}: A second EOM (EOM2) with the driving frequency of 6.856 GHz, producing a horizontal, linear polarized beam and simultaneous resonant transitions of $F=2\leftrightarrow F'=2$ and $F=1\leftrightarrow F'=2$, pumps the atoms into magnetic insensitive state $F=2,\,m_F=0$, Figure~\ref{subfig:FrequencyTransitions}. The atoms subsequently undergo two velocity selection + blow-away sequences: A first Raman $\pi-$pulse transfers the atoms from the $F=2$ to $F=1$ state, followed by a resonant light (produced by an un-modulated laser light) that blows away atoms that are remaining at the $F=2$ state. A second Raman $\pi-$pulse transfers the atoms back to $F=2$ state, and followed by a resonant light (produced by +1 sideband of EOM2, with driving frequency shifted to 6.869 GHz to avoid coupling of the carrier to atomic transitions) that blows away atoms that are remaining at $F=1$ state. This results in a narrow velocity spread of the atoms of $\approx 70$ nK in the vertical axis. Approximately 85 \% of the atoms are lost through the velocity-selection sequences, this corresponds to $\approx 1\times 10^7$ atoms for the interferometer. 
    \item \textbf{Interferometry}: A matter-wave Mach-Zehnder interferometry of $\frac{\pi}{2}-T-\pi-T-\frac{\pi}{2}$ is performed on the atomic fountain, with $T$ as long as 200 ms. EOM1's driving frequency is shifted to 6.58 GHz, and its driving amplitude is increased such that the +1 sideband to the carrier intensity ratio is $>$ 40\%. The driving frequency and amplitude are maintained throughout the interferometry sequence, producing a fixed Raman laser $\omega_1$. The chirping of the Raman laser ($\omega_2$) is then provided by linear sweeping a double-pass AOM via AD9854. The intensity ratio between $\omega_1$ and $\omega_2$ is varied via a half waveplate and polarizing beamsplitter (PBS) pair to minimize the differential AC Stark shift. A further mitigation of systematic errors due to the AC Stark shifts and residual magnetic field can be achieved by alternating the sign of the frequency chirping during every alternative measurement cycle.  
    \item \textbf{State-selective detection}: The falling atomic cloud first goes through a horizontal (circular-polarized) retro-reflecting light sheet near resonance at $F=2\leftrightarrow F'=3$, emitting fluorescent light proportional to the population of atoms at the $F=2$ state; most of the atoms at the $F=2$ state are lost due to thermal expansion from the imaging lights. The remaining falling atoms go through a second light sheet (vertically separated by 22 mm) having the same frequency but with a small portion of repumping light provided by EOM1, the atoms then emit fluorescent light proportional to the population of atoms at $F=1$ state, thus enabling state-selective detection. 
\end{enumerate}

\section{Performance}

\subsection{Vibration compensation through Raman lasers' phase offset}
\label{subsec:phaseCompensation}
\begin{figure}%[h]
%\centering
\includegraphics[width=0.48\textwidth]{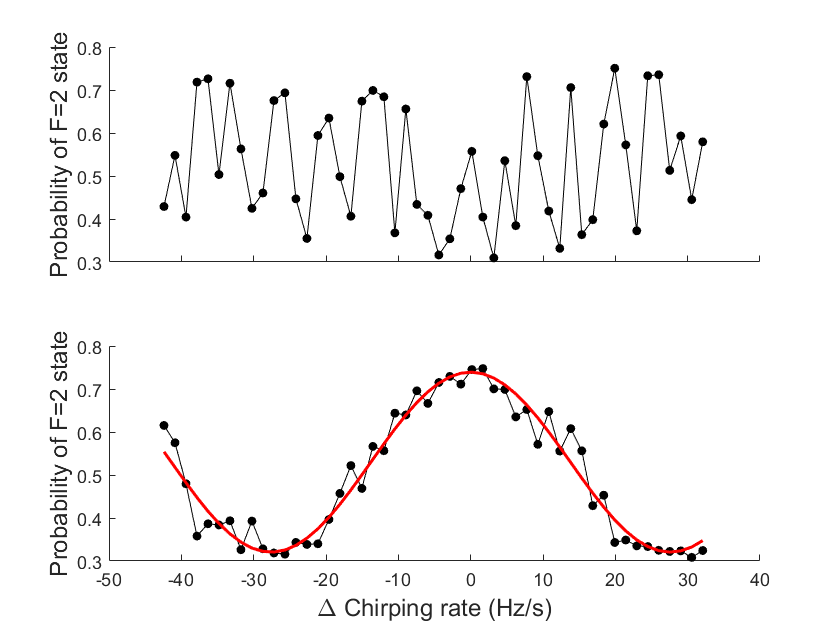} 
\caption[]{Center fringe scanning with 50 points frequency chirping steps, each step is equal to 1.52 Hz/s. `0' $\Delta$ chirping rate represents Raman frequency chirping rate at 25.0707355 MHz/s. Each point represents a single measurement of 2 s. The frequency of the artificial excitation is set at 0.95 Hz to avoid aliasing effect on the measurements. (Top) The interferometric fringe is washed out without phase compensation, (Bottom) interferometric fringe is restored with  Raman lasers' phase compensation. Red curve is the sine fit on the center fringe where the fringe's maxima represents frequency induced phase shift equal to the gravitational induced phase shift, $2\pi\Delta f T=k_\text{eff}gT^2$.} 
\label{fig:phasecancellation} 
\end{figure}

Vibration noise remains one of the pressing challenges for $\mu\text{Gal}$ precision transportable absolute gravimeters in noisy environments. In our experiment, the Raman lasers are retro-reflected vertically by a mirror. The mirror serves as a ``gravity reference'', as any vibrations incurred on the mirror will inevitably imprint on the interferometer. The vibrations of the retro-reflecting mirror have to be either actively cancelled~\cite{hensley1999active, oon2022compact} or by means of post processing~\cite{le2008limits} in order to minimize the inertial influence on the gravitational acceleration measurements. 
\par 
To demonstrate the ability of mitigating the effect of vibrations through Raman lasers' phase offset, we create an artificial excitation on the passive isolation platform (where the Raman lasers' retro-reflecting mirror is placed) through voice coil actuators (please refer to our previous work on the detail of the vibration isolation setup~\cite{oon2022compact}). At Raman separation time of $T=140$ ms, the presence of excitation at 0.95 Hz with amplitude of $10^{-5}$ g would render the interferometric fringe unrecognizable during Raman lasers' frequency scanning, as shown in Figure~\ref{fig:phasecancellation} (top). We then go through the same frequency scanning, but the vibration signals are recorded via an accelerometer (Nanometrics, Titan) during the three-pulse Mach-Zehnder interferometry, and the vibration induced phase shift ($\phi$) is then compensated at the third Raman's pulse via the recursive function, 
\begin{equation}
    \phi_\text{k+1}=\phi_\text{k}+k_\text{eff}\cdot f_k\cdot a_k\cdot dt,
\end{equation}
where $a_k$ is the acceleration of the retro-reflecting mirror at $k$-th iterative loop, $dt=10$ $\mu\text{s}$ sampling time and $f_k$ is the triangular shape transfer function symmetric around the end of second Raman pulse (t=0)~\cite{lautier2014hybridizing,geiger2011detecting} with $f_k=k\cdot dt\,(t<0)$ and $f_k=2T-k\cdot dt\,(t>0)$. Interferometric fringe is restored by feeding the $\phi_k$ output into the offset of the optical phase-locked loop, Figure~\ref{fig:phasecancellation} (bottom). A sine fit to the center fringe shows a resolution of 12.4 $\mu\text{Gal}$ ($\approx 1.26\times 10^{-8}$ g) after an integration time of 100 s, despite under the presence of a $10^{-5}$ g vibration excitation.

\subsection{Gravity measurements} 

\begin{figure*}[t!]
%\centering
\includegraphics[width=0.98\textwidth]{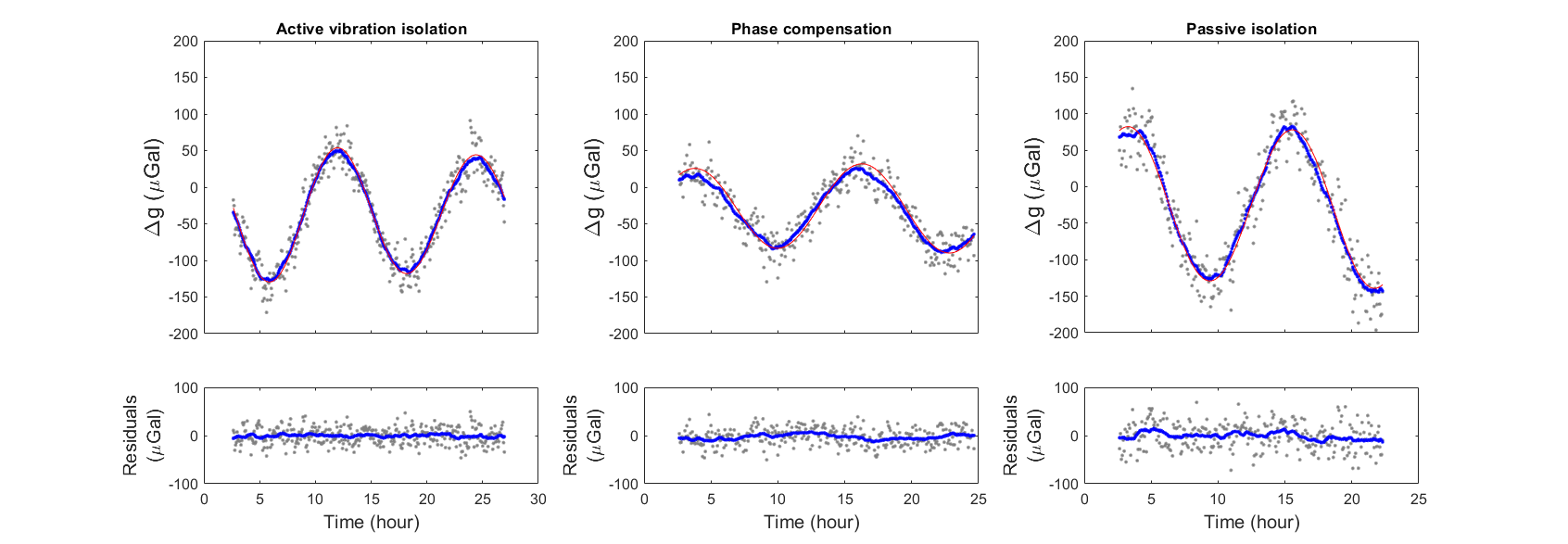} 
\caption[]{ (Top) Continuous 24-hour gravity measurements under three different scenarios (from left to right): active vibration isolation via voice coil actuators, Raman lasers' phase compensation (sitting on passive isolation stage) and with only passive isolation. Each gray point represents a single center fringe scanning (50 frequency steps, 250 s) and each blue point represents 29 points running average equivalent to $\approx$ 2 hours of averaging time. Red curves represent the Earth's tide model (micro-G Lacoste, Quick-Tide Pro) of the local gravitational variations at the corresponding coordinates and time. (Bottom) Residuals of gravity subtracted from Earth’s tide model.} 
\label{fig:tidal} 
\end{figure*}

\begin{table}[h]
\begin{center}
    \begin{tabular}{ | c | c | c |}
    \hline
Averaging time & 250 s & 2 h
    \\ \hline
Active vibration isolation  & 17.8 $\mu\text{Gal}$  & 2.5 $\mu\text{Gal}$
    \\ \hline
Phase compensation & 18.0 $\mu\text{Gal}$ & 5.1 $\mu\text{Gal}$
    \\ \hline
Passive isolation & 27.5 $\mu\text{Gal}$ & 7.0 $\mu\text{Gal}$
    \\ \hline
\end{tabular}
\end{center}
\caption[]{Measurement stability under active vibration isolation, phase compensation and passive isolation.}
\label{Table:sDeviations}
\end{table}

Figure~\ref{fig:tidal} shows continuous gravity measurements under three different scenarios, namely, (1) with active vibration isolation where the Raman's retro-reflecting mirror is actively stabilized by voice coil actuators~\cite{oon2022compact}, (2) vibration compensation via Raman lasers' phase offset and (3) with only passive vibration isolation (Minus K Technology, 25BM-10). The standard deviations of the three gravity measurements with both averaging time of 250 s and 2 h are shown in Table~\ref{Table:sDeviations}.

\begin{figure}%[h]
%\centering
\includegraphics[width=0.48\textwidth]{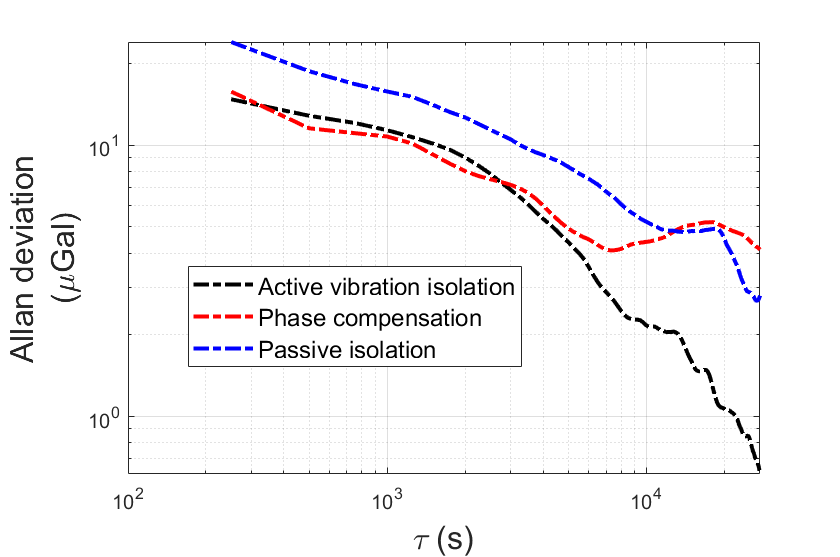} 
\caption[]{Allan deviations of the gravity measurements. (Black) Resolution of 1 $\mu\text{Gal}$ with $\approx 6$ hours integration time under active vibration isolation. (Red) Raman lasers' phase compensation with hour-timescale oscillation and (Blue) measurements under passive vibration isolation only.} 
\label{fig:allanD} 
\end{figure}

 It is worth noting that the temperature of the accelerometer (Nanometrics, Titan) has to be well stabilized (within $\pm 0.01^{\circ}$ C) to avoid feedback of the thermal noise signals onto the gravity measurements. To further reduce the thermal noise, a 0.001 Hz digital high-pass filter is employed on the accelerometer signal output when implementing active vibration isolation (a 0.008 Hz high-pass filter when implementing the Raman lasers' phase compensation). Vibration cancelation using the Raman lasers' phase compensation is prone to the influence of the accelerometer signal bias; the integration of the acceleration signals drifting from zero volt can render significant oscillations on the gravitational measurements, as evidenced by the Allan deviations of the measurements, Figure~\ref{fig:allanD}. Further reduction of such an effect will be addressed in future work~\cite{le2008limits}. Active vibration isolation is less sensitive to the slow frequency feedback compared to using phase compensation, this is due to the fact that the frequency response of the passive isolation stage to the force produced by the voice coil actuator has an additional high-pass characteristic at low frequency ($1/\omega$ for $\omega<\omega_0$ and $1/\omega^2$ for $\omega<<\omega_0$, where $\omega_0$ is the resonant frequency of the passive isolation stage )~\cite{hensley1999active, en2021transportable, freier2010measurement}.

\section{Conclusions}
We have demonstrated a compact single-seed laser system on a transportable high-precision $^{87}$Rb-based atomic gravimeter that reaches measurement stability of 2.5 $\mu\text{Gal}$ with averaging time of 2 h. The optical phase-locked loop derived from a single-seed laser provides a convenient way for Raman laser frequency chirping and vibration compensation through Raman lasers' phase offset. Frequency generation using a mixture of an AOM and a EOM helps in alleviating detrimental effects on the accuracy of the gravity measurements due to parasite Raman transitions. The modular design of the optical systems ensures its mechanical stability during transportation as well as plug-and-play capability of the individual optic modules (provided that the same optical fiber is used). All these features offer a valuable approach for the construction of future in-field atomic gravimeters as well as other atomic technologies.

\begin{acknowledgments}
The authors thank Christoph Hufnagel, Nathan Shetell, Kaisheng Lee and Koon Siang Gan for their careful reading of their manuscript and their many insightful comments and suggestions. This work is supported by the National Research Foundation (Singapore) under the Quantum Engineering Program, DSO National Laboratories (Singapore) and the Ministry of Education of Singapore.
\end{acknowledgments}

\section*{AUTHOR DECLARATIONS}
\subsection*{Conflict of interest}
The authors have no conflicts to disclose.

\section*{DATA AVAILABILITY}
The data that support the findings of this study are available from the corresponding author upon reasonable request.

%\clearpage

\bibliography{biblio}% Produces the bibliography via BibTeX.

\end{document}